\documentclass[aps,prl,twocolumn,superscriptaddress]{revtex4}

\usepackage{graphicx}
\usepackage{color}
\usepackage{amsfonts}
\begin{document}
\title{Electron-phonon coupling in quasi free-standing graphene}

\author{Jens Christian Johannsen}
\affiliation{Department of Physics and Astronomy, Interdisciplinary Nanoscience Center, Aarhus University,
8000 Aarhus C, Denmark}
\author{S\o ren~Ulstrup}
\affiliation{Department of Physics and Astronomy, Interdisciplinary Nanoscience Center, Aarhus University,
8000 Aarhus C, Denmark}
\author{Marco~Bianchi}
\affiliation{Department of Physics and Astronomy, Interdisciplinary Nanoscience Center, Aarhus University,
8000 Aarhus C, Denmark}
\author{Richard~Hatch}
\affiliation{Department of Physics and Astronomy, Interdisciplinary Nanoscience Center, Aarhus University,
8000 Aarhus C, Denmark}
\author{Dandan~Guan}
\affiliation{Department of Physics and Astronomy, Interdisciplinary Nanoscience Center, Aarhus University,
8000 Aarhus C, Denmark}
\author{Federico Mazzola}
\affiliation{Department of Physics and Astronomy, Interdisciplinary Nanoscience Center, Aarhus University,
8000 Aarhus C, Denmark}
\author{Liv Hornek\ae r}
\affiliation{Department of Physics and Astronomy, Interdisciplinary Nanoscience Center, Aarhus University,
8000 Aarhus C, Denmark}
 \author{Felix Fromm}
 \author{Christian Raidel}
 \author{Thomas Seyller}
 \affiliation{Institut f{\"u}r Physik der Kondensierten Materie, Universit{\"a}t Erlangen-N{\"u}rnberg, Erwin-Rommel-Strasse 1, D-91058 Erlangen, Germany}
\author{Philip~Hofmann}
\affiliation{Department of Physics and Astronomy, Interdisciplinary Nanoscience Center, Aarhus University,
8000 Aarhus C, Denmark}
\email[]{philip@phys.au.dk}

\date{\today}

\begin{abstract}
Quasi free-standing monolayer graphene can be produced by intercalating species like oxygen or hydrogen between epitaxial graphene and the substrate crystal. If the graphene is indeed decoupled from the substrate, one would expect the observation of a similar electronic dispersion and many-body effects, irrespective of the substrate and the material used to achieve the decoupling. Here we investigate the 
electron-phonon coupling in two different types of quasi free-standing monolayer graphene: decoupled from SiC via hydrogen intercalation and decoupled from Ir via oxygen intercalation. Both systems show a similar overall behaviour of the self-energy and a weak renormalization of the bands near the Fermi energy. The electron-phonon coupling is found to be sufficiently weak to make the precise determination of the coupling constant $\lambda$ through renormalization difficult. The estimated value of $\lambda$ is $0.05(3)$ for both systems.
\end{abstract}

\maketitle


The unusual electronic structure of graphene  \cite{Novoselov:2004,Novoselov:2005}  has significant implications on the transport properties as well as on the many-body effects in this material. An interaction of particular importance for applications is the electron-phonon coupling, as this mechanism can be dominant for the scattering of carriers near the Fermi surface. In theoretical work, it was recognised early that the Dirac-like dispersion and its corresponding restriction of scattering phase space renders the electron-phonon coupling in graphene quite different from that in ordinary metals \cite{Calandra:2007,Park:2007a}. In particular, it  leads to a doping-dependent electron-phonon coupling strength $\lambda$ and a much more complex shape of the self-energy than in ordinary metals, where the density of states near the Fermi level, and its contribution to the self-energy,  is approximately constant on a phonon energy scale. 

Angle-resolved photoemission (ARPES) is a particularly valuable technique to address many-body interactions experimentally since it provides detailed information on the energy and $k$-dependent lifetime of the carriers. Several ARPES studies of the electron-phonon coupling in graphene have been published \cite{Bostwick:2007,Bostwick:2007b,Zhou:2008b,Park:2008b,Bianchi:2010,McChesney:2010,Forti:2011,Pletikosic:2012}, mostly for strongly $n$-doped graphene  ($n \approx 10^{13}$~cm$^{-2}$), where the electron-phonon scattering was found to be of intermediate strength with  $\lambda$ in the order of $0.2 - 0.3$.

For weakly doped graphene, the reduction of scattering phase space implies a weaker and eventually vanishing electron-phonon coupling strength. Experimentally, it is  difficult to verify this with high precision because of the often employed methodology to determine $\lambda$. This relies on the observed energy dependence of the electronic self-energy near the Fermi energy $E_F$, where electron-phonon coupling gives rise to kinks in the dispersion and a  decrease of the linewidth. For very weak coupling, these spectral features are equally weak and hard to distinguish from the noise in the spectra. It should also be mentioned that the determination of $\lambda$ in this way only works for sample temperatures that are low compared to the typical excitation temperature for phonons. This is seemingly easy to assure for graphene with its high Debye temperature, but it has been questioned if the Debye temperature or the doping-dependent and lower Bloch-Gr\"uneisen temperature sets the relevant scale for electron-phonon coupling in graphene \cite{Efetov:2010}. Nevertheless, there is mounting experimental evidence that the electron-phonon coupling in weakly doped graphene is indeed quite small \cite{Siegel:2011,Forti:2011,Ulstrup:ARXIV}. 

An essential question for any meaningful comparison between experimentally determined many-body effects and calculations is of course how much the properties of the epitaxial graphene, typically used in experiments, actually resemble those of the pristine graphene assumed in calculations. Evidence from transport measurements suggest that the contact of graphene to a substrate can significantly reduce the carrier mobility, especially at higher temperatures \cite{Morozov:2008,Chen:2008}. Epitaxial graphene on both SiC and transition metal surfaces still exhibits a relatively strong interaction with the substrate, at least strong enough for the graphene to have a well-defined azimuthal orientation with respect to the substrate. In rare cases, such as for Pt(111) \cite{Preobrajenski:2008b}, this well-defined azimuthal orientation is lost, but for the majority of samples even stronger interactions are directly observed in the dispersion of the graphene bands, ranging from weak replica bands with mini-gaps on Ir(111) \cite{Kralj:2011,Pletikosic:2009} to interactions which are so strong that the typical graphene dispersion is not observed, even though the structure of the layer is graphene-like \cite{Emtsev:2008,Sutter:2010}. Apart from these largely structural effects, a relatively strong doping is also frequently observed. Note that even in the absence of a graphene-substrate interaction immediately evident in the band dispersion, the different dielectric properties of the substrate (wide-gap semiconductor vs. metal) might still affect the many-body interactions in graphene. 

In this paper, we investigate the so-called quasi free-standing monolayer graphene (QFMLG) as a possible experimental solution to the problem of the graphene-substrate interaction.  QFMLG was first proposed in connection with hydrogen-intercalation below the graphene-like so-called buffer layer forming the $(6\sqrt{3} \times 6\sqrt{3})$R$30^{\circ}$ structure on SiC(0001) \cite{Riedl:2009}. Whereas the buffer layer does not show a graphene-like electronic structure \cite{Emtsev:2008}, a weakly $p$-doped and narrow Dirac cone emerges upon hydrogen intercalation \cite{Bostwick:2010}.  Similar results can be achieved by oxygen intercalation for graphene on Ir(111) \cite{Larciprete:tbp} where the graphene is also $p$-doped but any other indication of graphene-substrate interaction is lost. Here we probe the electron-phonon coupling on these two QFMLG systems that have been synthesised via hydrogen intercalation on SiC(0001) and oxygen-intercalation on Ir(111). We find that the electron-phonon coupling is weak $\lambda \lesssim 0.07$ for both systems. In view of the rather different substrates and materials used for intercalation, this is taken as a suggestion that the QFMLG can indeed be taken as a model for pristine graphene. 

High-quality graphene was prepared  using standard methods on Ir(111) \cite{Coraux:2009}. The sample quality was checked with low-energy electron diffraction, showing an intense moir\ifmmode \acute{e}\else \'{e}\fi{} pattern for a clean graphene monolayer. Intercalation of oxygen was achieved by placing the sample within a custom made high-pressure O$_2$ doser and maintaining a background O$_2$ pressure at $4\times10^{-4}$~mbar for 10 min while keeping the sample temperature at 523~K. H-intercalated QFMLG on SiC(0001) was prepared \emph{ex-situ} by the methodology outlined in Ref. \cite{Speck:2011}. In the ARPES vacuum chamber, these latter samples were cleaned by annealing to 673~K. 
 The ARPES measurements were carried out at the SGM-3 beamline of the synchrotron radiation source ASTRID \cite{Hoffmann:2004} under ultrahigh vacuum conditions with a base pressure in the 10$^{-10}$~mbar range and with the sample temperature kept at 70~K. The employed photon energies were 47~eV and  32~eV for graphene on Ir and SiC, respectively, and the total energy and $k$ resolution were better than 18~meV and 0.01~\AA$^{-1}$, respectively.

\begin{figure}
\begin{center}
\includegraphics[width=1 \columnwidth]{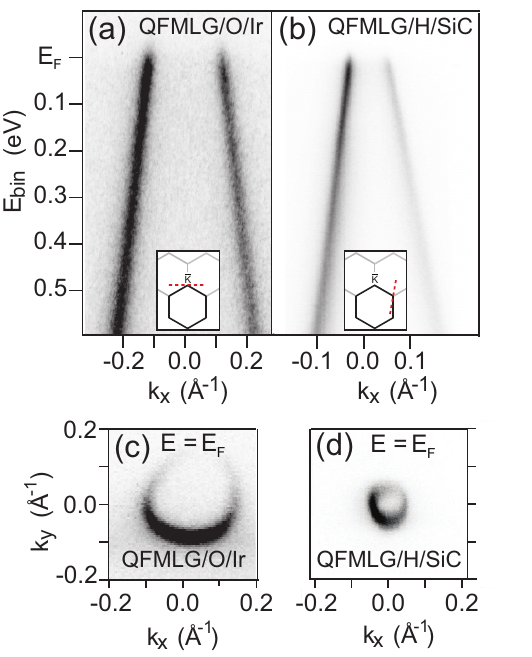}
\caption{Electronic structure of QFMLG on Ir(111) (a,c) and SiC(0001) (b,d) measured by ARPES. Shown is the photoemission intensity as a function of $k$ and binding energy (on the top part) and as a function of both components of $k$ at the Fermi energy (on the bottom part). Dark corresponds to high intensity. The insets show the direction of the shown dispersion within the Brillouin zone.
\label{fig:1}}
\end{center}
\end{figure}

The measured spectral function for QFMLG on both substrates is shown in Figure \ref{fig:1}. The data are cuts through a three-dimensional data set (taken as a function of electron kinetic energy and two emission angles) close to the Fermi energy $E_F$ and near the $\bar{K}$-point of the Brillouin zone. Note that here the  $\bar{K}$-point is defining the origin of the $k$-coordinate system. The characteristic Dirac cone is easily identified. The observed electronic structure is drastically different from the situation without intercalated oxygen or hydrogen: On Ir(111) the replica bands are not observed any more and on SiC(0001) the intercalation is required for any observable Dirac cone. In both cases, we find a clear $p$ doping with estimated carrier densities of $4\times 10^{13}$~cm$^{-2}$~(Ir) and $5\times 10^{12}$~cm$^{-2}$~(SiC). On Ir(111), the $p$ doping is ascribed to a charge transfer to the intercalated oxygen. On SiC(0001), it is caused by a spontaneous polarization of the substrate \cite{Ristein:ARXIV}. 

Data as in Figure \ref{fig:1} form the basis for a more detailed analysis of the electron-phonon coupling. For QFMLG on Ir(111), this analysis is carried out using the dispersion perpendicular to the $\bar{\Gamma}-\bar{K}$ direction. For QFMLG on SiC, the chosen direction is not a high-symmetry direction but encloses an angle of 40.4$^{\circ}$ with the  $\bar{\Gamma}-\bar{K}$ direction. Carrying out the same analysis for other directions in $k$-space does not lead to a significantly different result, consistent with the theoretical expectation of a uniform electron-phonon coupling strength over the Fermi contour \cite{Park:2008b}. 

Information about the electron-phonon coupling needs to be extracted in two steps. The first is to obtain the self-energy from the data and the second to extract information about the many-body effects from the self-energy. Assuming no $k$-dependence of the self-energy $\Sigma$ and no strong matrix element effects, ARPES measures the spectral function 
\begin{equation}
 \mathcal{A}(\omega,k)=\frac{\pi^{-1}|\Sigma''(\omega)|}{[\hbar \omega-\epsilon({k})-\Sigma'(\omega)]^{2}+\Sigma''(\omega)^{2}},
\label{equ:1}
\end{equation}
where $\hbar \omega$ is the binding energy, $\epsilon(k)$ the bare dispersion (without many-body effects) and $\Sigma$ the self-energy with real part $\Sigma'$ and imaginary part $\Sigma''$. For a cut at constant energy, a so-called momentum distribution curve (MDC), this has the form of a Lorentzian line with the width at a certain $\omega$ given by the slope of the band and $\Sigma''$, and the position by $\epsilon(k) + \Sigma'(\omega)$. The challenge for the analysis is that neither $\Sigma$ nor  $\epsilon(k)$ is known. Nevertheless, it is often possible to extract these quantities by using only a few extra assumptions, such as particle-hole symmetry or the fact that $\Sigma'$  and $\Sigma''$ are related by a Kramers-Kronig relation. 

\begin{figure}
\begin{center}
\includegraphics[width=1.0\columnwidth]{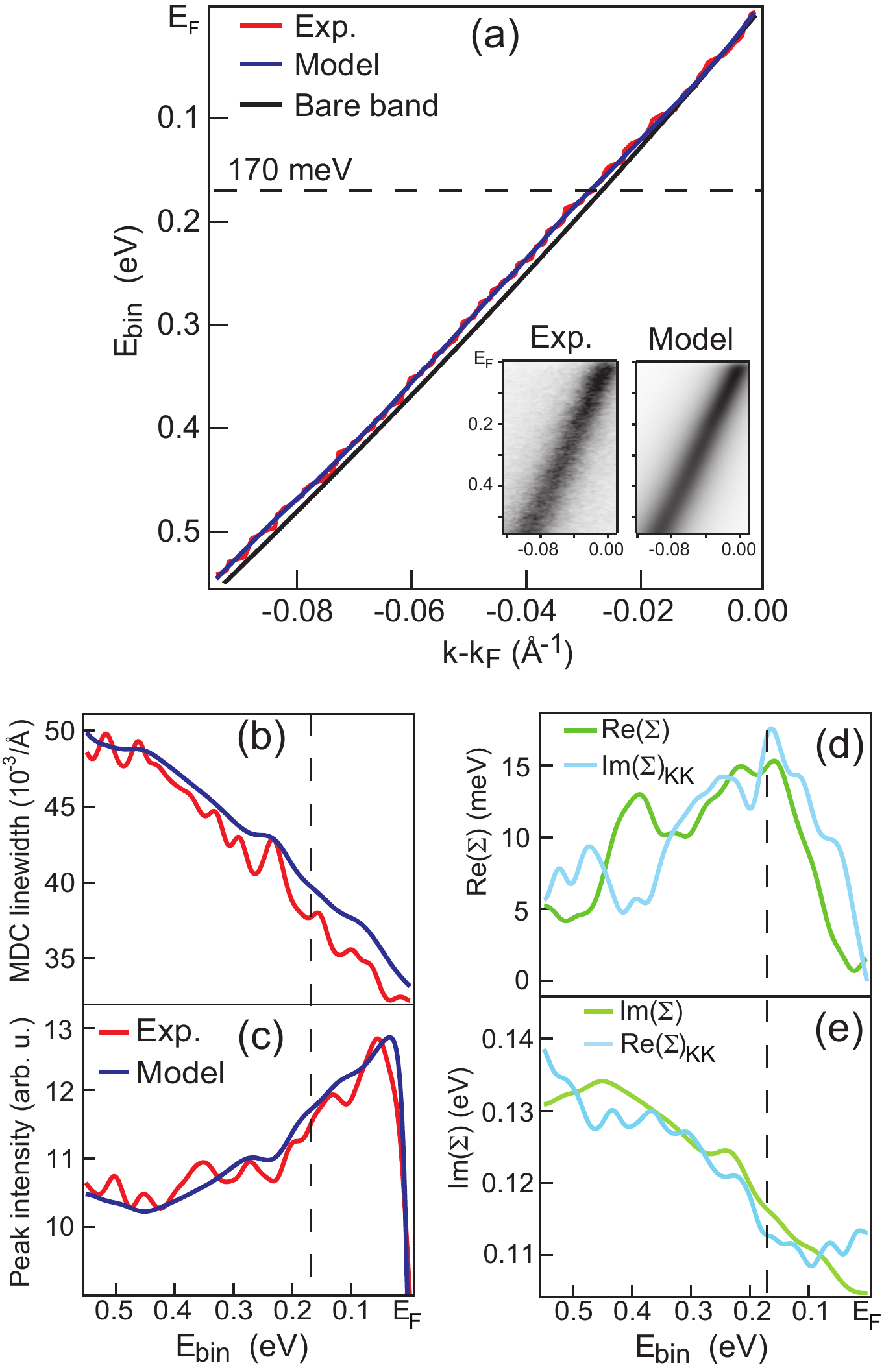}
\caption{Extraction of the self-energy $\Sigma$ for QFMLG on Ir(111).(a) Measured dispersion (MDC maximum) together with the model dispersion and the extracted bare dispersion. The inset shows a close-up of the experimental spectral function close to the Fermi level and the spectral function calculated from the derived bare dispersion and self-energy.  (b) Measured MDC width together with modelled MDC width. (c) Measured amplitude of the spectral function together with modelled amplitude. (d) and  (e): Demonstration of self-consistency, showing the modelled real part of the self-energy together with a Kramers-Kronig transformation (KK) of the imaginary part and vice versa. 
\label{fig:2}}
\end{center}
\end{figure}

Here we use a particularly efficient iterative approach to the problem that was very recently introduced by Pletikosi\ifmmode \acute{c}\else \'{c}\fi{} \emph{et al.} \cite{Pletikosic:2012}. For obtaining the self-energy and amplitude, the data have been smoothed by downsampling the number of energy points.
Figure \ref{fig:2} gives the detailed result of this method as applied to QFMLG on Ir(111). The procedure determines the self-energy and the bare dispersion and these can be used to model the spectral function. The result of the model can be compared to the experimental data. 
Figure \ref{fig:2} shows this comparison as well as a demonstration of the results' self-consistency. 

Figure \ref{fig:2}(a) shows the MDC peak position extracted from the data together with the peak position obtained from the model and the bare dispersion. We find an excellent agreement between experimental and modelled peak positions. It is sufficient to model the bare dispersion by a second order polynomial. The obtained curvature is positive, as expected for the $\pi$ band of graphene below the Dirac point. At a binding energy of $\approx$170~meV, the bare dispersion starts to deviate from the measured and modelled dispersion. This is the indication of a kink that is caused by electron-phonon coupling. The kink is also visible in the raw spectral function, but only barely. 

Figure \ref{fig:2}(b) shows the energy-dependent MDC width obtained from the experimental data and from the model. The kink in the dispersion should correspond to a width increase at the same energy ($\approx$170~meV) but this is hard to discern in the data over the overall increase of the width in the energy range plotted here. In any event, the measured and modelled widths agree very well. 

According to (\ref{equ:1}), the self-energy also determines the amplitude of the spectral function and Figure \ref{fig:2}(c) shows a comparison between  experimental and modelled  peak intensity. Again, the agreement is excellent. The peak amplitude increases notably for energies close to the Fermi energy and this can also be seen in the raw data. The peak width of the MDCs near the Fermi energy ($\approx$0.032~\AA$^{-1}$) is comparable, if slightly larger, to the best values obtained for non-intercalated graphene on Ir(111) \cite{Kralj:2011,Pletikosic:2009}.

Finally, Figure \ref{fig:2}(d) and (e) show the real and imaginary part of the self-energy, respectively. Not only the extracted real (imaginary) part is shown, but also the Kramers-Kronig transformed imaginary (real) part, illustrating the self-consistency of the result. The real part of the self-energy $\Sigma'$ shows a peak at $\approx 170$~meV and so does the Kramers-Kronig transformation of the imaginary part. The corresponding weak step in $\Sigma''$ is only discernible in the Kramers-Kronig transformation of $\Sigma'$. Overall, a satisfactory degree of self-consistency is achieved. 

\begin{figure}
\begin{center}
\includegraphics[width=1.0\columnwidth]{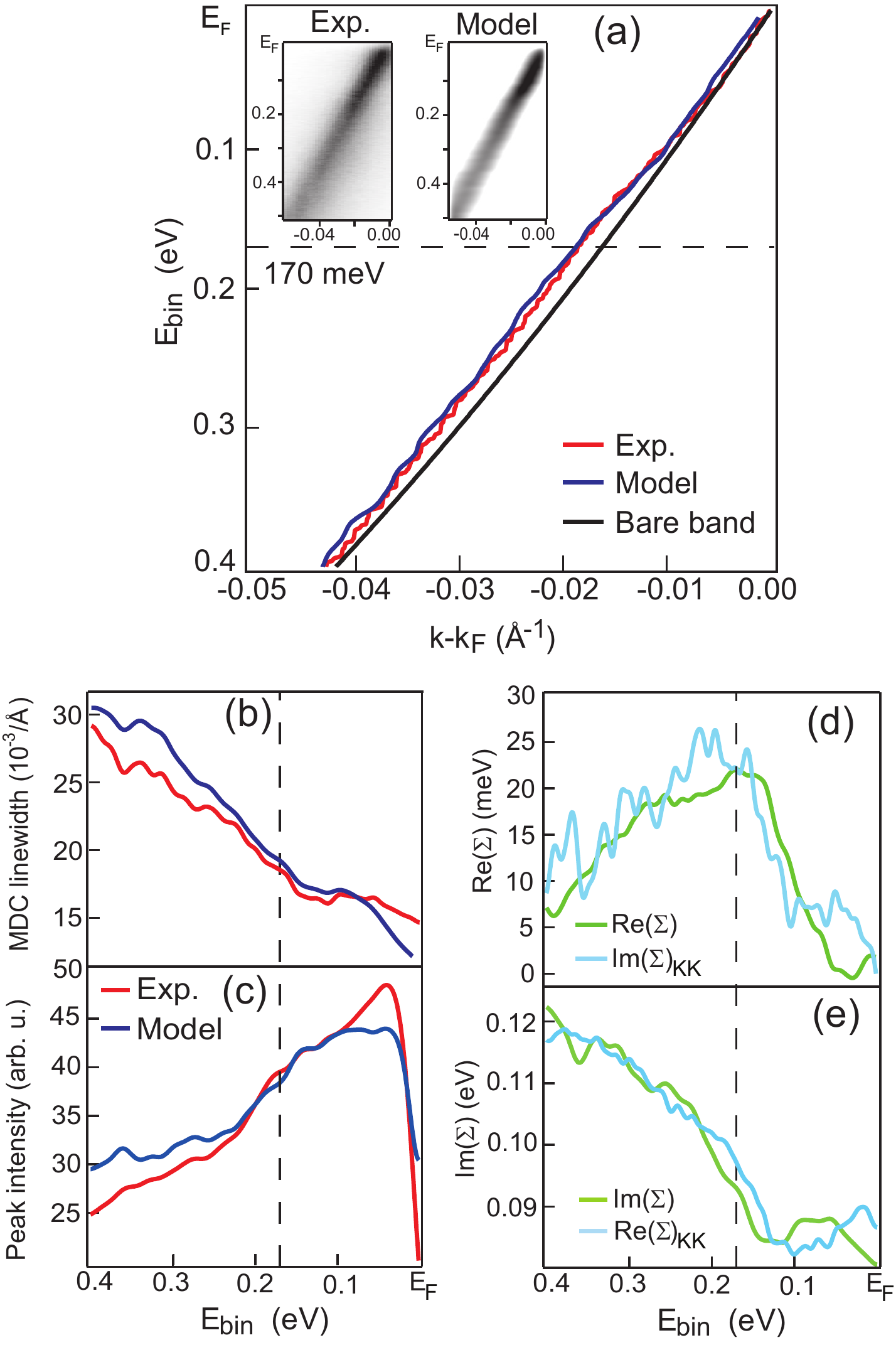}
\caption{Extraction of the self-energy $\Sigma$ for QFMLG on SiC(0001).(a) Measured dispersion (MDC maximum) together with the model dispersion and the extracted bare dispersion. The inset shows a close-up of the experimental spectral function close to the Fermi level and the spectral function calculated from the derived bare dispersion and self-energy.  (b) Measured MDC width together with modelled MDC width. (c) Measured amplitude of the spectral function together with modelled amplitude. (d) and  (e): Demonstration of self-consistency, showing the modelled real part of the self-energy together with a Kramers-Kronig transformation (KK) of the imaginary part and vice versa.
\label{fig:3}}
\end{center}
\end{figure}

The same analysis for QFMLG on SiC(0001) is shown in Fig. \ref{fig:3}. It gives very similar results. Again, the prominent structure in the data is a kink at $\approx 170$~meV that is also identified in the linewidth and in the peak intensity. The agreement between the data and the model, as well as the self-consistency is equally good as for the QFMLG on Ir(111). The data are quite similar to the recent results by Forti \emph{et al.} \cite{Forti:2011}.

The central result of this paper lies in the comparison of the results from the two QFMLG samples. Qualitatively, both show the theoretically expected behaviour for hole-doped graphene  \cite{Calandra:2007,Park:2007a}: in a region close to the Fermi energy, $\Sigma''$ is relatively constant, followed by a steep increase at $\approx 170$~meV and a linear increase at even higher energies. In a simple picture, this behaviour is explained as follows: for weakly-doped graphene, low energy acoustic phonons play an insignificant role for the electron-phonon coupling, as seen by the flat $\Sigma''$ near the Fermi level. The most significant coupling is to an optical phonon mode at 170~meV. The linear increase at even higher energies is also related to the electron-phonon coupling, but it stems from an increase in the electronic density of states away from the Dirac point, an effect that can have a pronounced influence on the electron-phonon coupling strength in semimetals \cite{Gayone:2003}. 

The dominant role of the optical phonons in the scattering process is consistent with all other observations, such as the kink in the real part of the self-energy as well as the narrowing of the MDC peak and its increased intensity for binding energies smaller that 170~meV.  The dominant coupling to optical phonons is also consistent with several other reported observations for electron-phonon coupling in graphene \cite{Bostwick:2007,Bostwick:2007b,Zhou:2008b,Park:2008b,Bianchi:2010,McChesney:2010,Forti:2011,Pletikosic:2012}. 

Another effect that contributes to the self-energy is the electron-electron interaction. It can be argued, however, that this effect is only of minor importance for the energy region and the kink studied here. Indeed, electron-electron interaction effects such as electron-hole and electron-plasmon processes have been shown to depend critically on the position of the Dirac point with respect to the Fermi level and thus on the doping \cite{Bostwick:2007,Bostwick:2010}. 
The observed band renormalization reported here occurs at a fixed binding energy of $\approx 170$~meV regardless of doping, which is seen by comparing the dispersions for the two substrates in Figs. \ref{fig:2}(a) and \ref{fig:3}(a). Additionally, since the most dominant optical phonon modes are found at this energy in calculations \cite{Calandra:2007,Park:2007a,Vitali:2004}, we do not consider interactions associated with electrons to significantly influence the observed kink.

Note that there is an absolute offset between the different MDC linewidth and $\Sigma''$ values for the two samples near the Fermi energy. The conventional interpretation of this would be a smaller amount of defect scattering for QFMLG on SiC(0001) than on Ir(111). This does not necessarily have to be the case, however, because the role of not completely intercalated graphene regions would be quite different for the two systems. On Ir(111), areas that are not completely (or not at all) intercalated with oxygen would still give rise to a Dirac cone, only at a different energy and this would broaden the spectra. On SiC, on the other hand, areas without hydrogen intercalation would not at all contribute to the observed Dirac cone. 

From the self-energy, we can now attempt a determination of the electron-phonon mass enhancement parameter $\lambda$. As mentioned above, the key-problem with this type of analysis is that the determination of $\lambda$ becomes quite uncertain in the case of weak coupling, simply because it is difficult to quantify an effect that is weak enough to be hardly observable. A far superior approach to determine $\lambda$ in such a case is via a study of the temperature-dependent linewidth. For temperatures at or above the Debye temperature, this linewidth is directly proportional to $\lambda$, permitting a straight-forward determination of the scattering strength \cite{McDougall:1995,Gayone:2003,Kim:2005a,Hofmann:2009b,Ulstrup:ARXIV}. Unfortunately, this is not an option for QFMLG because the system already becomes unstable for temperatures far below the Debye temperature. 

We thus restrict ourselves to a simple order of magnitude estimate  of $\lambda$, considering only the phonon mode at 170~meV and relating the step in $\Sigma''$ to $\lambda$ via \cite{Fink:2006}
\begin{equation}
\lambda=\frac{2}{\pi} \frac{\Delta \Sigma''}{\hbar \omega_{ph}}, 
\end{equation}
where $\Delta \Sigma''$ is the size of the jump in $\Sigma''$ and $\hbar \omega_{ph}$ is the phonon energy. While it is difficult to obtain a precise value for $\Delta \Sigma''$, an estimate consistent with both samples is 10~meV$  \lesssim \Delta \Sigma'' \lesssim 20$~meV. This leads us to estimate a value of $\lambda = 0.05(3)$.  It is also possible, in principle, to determine $\lambda$ from the slope of the real part of the self-energy. This determination is consistent with that from $\Delta \Sigma''$ but the uncertainties are even larger. 

A value of $\lambda = 0.05(3)$ compares well to the few other studies of weakly doped graphene. Forti  \emph{et al.} have published a study for QFMLG on SiC(0001) and analyzed the MDC linewidth using a somewhat different approach from the one employed here. They found $\lambda \approx 0.03$, a value consistent with ours when taking into account the large uncertainties of both experiments. Ulstrup \emph{et al.} have studied epitaxial graphene on Ir(111), i.e. without oxygen intercalation, via temperature-dependent ARPES and found $\lambda \approx 9 \times 10^{-4}$ \cite{Ulstrup:ARXIV}. 

%

The most important conclusion from this work is that quasi free-standing monolayer graphene does indeed appear to be quasi free-standing in terms of the observed electron-phonon coupling: while the systems studied here are very different in terms of substrate materials, a wide-gap semiconductor vs. a metal, and intercalation materials (H, O), the resulting electron-phonon coupling is similar in strength and spectral appearance. Another important result is that the coupling is small, consistent with what would be expected for pristine graphene. 

This work was supported by The Danish Council for Independent Research / Technology and Production Sciences and the Lundbeck foundation. The authors express their gratitude towards  Silvano~Lizzit, Rosanna Larciprete, Paolo Lacovig, Matteo Dalmiglio, Fabrizio Orlando and Alessandro Baraldi for sharing the technique of oxygen intercalation under graphene on Ir(111) prior to publication.


\end{document}